\documentclass[review,3p,twocolumn,number,sort&compress]{elsarticle}

\usepackage[T1]{fontenc}
\usepackage[utf8]{inputenc}
\usepackage{natbib}
\usepackage[english]{babel} 
\usepackage{amsmath,amsfonts,amsthm}
\usepackage{amssymb}
\usepackage{graphicx} 
\graphicspath{{./figures/}}
\usepackage{float}
\usepackage{booktabs}
\usepackage{siunitx}
\usepackage{url}
\usepackage{hyperref}
\usepackage[noabbrev]{cleveref}
\usepackage{lmodern}
\usepackage{caption}
\usepackage{verbatim}
\usepackage{ragged2e}
\usepackage{flushend}
\usepackage{textcomp}
\usepackage{breakcites}
\usepackage{microtype}
\usepackage{tabularx}
\usepackage[toc,page]{appendix}
\usepackage{pdflscape}
\usepackage{ctable}
\usepackage{afterpage}
\usepackage{subcaption}

\journal{Nuclear Physics A}

\begin{document}

\begin{frontmatter}



\title{The XBPF, a new multipurpose scintillating fibre monitor for the measurement of secondary beams at CERN}


\author{I. Ortega Ruiz}
\ead{inaki.ortega@cern.ch}
\author{L. Fosse, J. Franchi, A. Frassier, J. Fullerton, J. Kral, J. Lauener, T. Schneider, G. Tranquille}

\address{CERN, 1211 Geneva, Switzerland}


\begin{abstract}
  The Beam Instrumentation group at CERN has developed a new scintillating fibre beam monitor for the measurement of secondary particle beams in the CERN Experimental Areas.
The monitor has a simple design that stands out for its low material budget, vacuum compatibility, good performance, low cost, and ease of production.
By using different read-out techniques the monitor can perform several functions, such as measurement of the profile, position and intensity of the beam, momentum spectrometry, generation of fast trigger signals, and measurement of the time-of-flight for particle identification.
The monitor has been successfully commissioned in the recently created test beams of the CERN Neutrino Platform, where it has shown an excellent performance as described in the paper.
\end{abstract}


\end{frontmatter}


\section{Introduction} {\label{sec:introduction}}
The Experimental Areas at CERN host a rich and diverse program of high-energy physics experiments and research and development in particle detectors and accelerator technology.
These facilities provide on user's request a rich variety of secondary beams ($e^{-/+}, p^{+/-}, \pi^{+/-}, \mu^{+/-}, K^{+/-}$, and diverse heavy ions) over a wide range of energies and intensities, typically \SIrange{0.1}{450}{GeV} and $10^2$ to $10^7$ particles per second per \si{mm^2} \cite{sba_web}.
The transverse profile and position of these beams is measured with various types of wire gaseous chambers, mainly Multi-Wire Proportional Chambers (MWPC) and Delay Wire Chambers (DWC) \cite{dreesen1978integrating} that date from the late 70's.
The performance of many of these monitors is seriously compromised due to ageing problems and radiation damage and their maintenance is very difficult due to outdated components and loss of expertise.
Furthermore, these wire chambers cannot fulfil the requirements of a new test beam facility at CERN, the Neutrino Platform, which has further motivated the search for a new technology for their replacement.

The CERN Neutrino Platform has been created in the framework of an international collaboration on R\&D for neutrino detection technologies.
This new facility is instrumented with two newly constructed beam lines -- H2-VLE and H4-VLE described in \cite{charitonidis2017low} -- that provide low energy and low intensity secondary beams of \SIrange{0.3}{12}{GeV/c} in bursts of 10 to 1,000 particles per burst.
The duration of a burst is typically \SI{4.8}{s} and the average beam spot size is large, in the order of several \si{cm} diameter.
The first experiments using this facility are the two ProtoDUNE detectors -- NP02 and NP04 -- which serve to study the detection technologies that will be used later by the DUNE Experiment in the United States \cite{manenti2017protodunes,acciarri2016long}.

Due to the special characteristics of these beam lines, they require beam monitors with a detection efficiency higher than 90\%, space resolution per channel of at most \SI{1}{mm}, active area of $\sim\SI{20 x 20}{cm}$, and very low material budget to avoid an emittance growth of the low energy beams.
Additionally, the beam instrumentation was requested to provide the following functionalities, which are critical for the successful characterisation of the ProtoDUNE detectors: single-particle tracking, momentum spectrometry, particle identification via Time-Of-Flight (TOF), and generation of fast trigger signals.
All events recorded by the beam instrumentation must be also time stamped with a clock distributed over White Rabbit (WR), which is an Ethernet-based network for sub-nanosecond accuracy timing distribution \cite{wr}.
WR is used to establish a common time reference among the beam instrumentation and the ProtoDUNE Experiments, in such a way that the information from all systems can be precisely correlated in time.
Further information on the characteristics of these beam lines and their requirements in terms of instrumentation can be found in \cite{h2,h4}.

The Beam Instrumentation group at CERN has been investigating a replacement for the ageing MWPC and DWC that has led to identifying scintillating fibres (SciFi) as the best suited detection technique, as reported in \cite{ortega2017jacow,ortega2018accurate,ortega2019multipurpose}.
It offers several advantages, such as a low material budget, good performance, easy scalability, and the low cost of scintillating fibres.
Additionally, the fibres can be easily integrated in the vacuum of the CERN Experimental Areas, further minimising the material budget by avoiding vacuum windows and interruptions of the beam line.
A comparison of the material budget of scintillating fibres with the current gaseous detectors is shown in \cref{tab:results_material}.
More details about this calculation can be found in \cite{ortega2018accurate} (Appendix D.1, page 163).

\begin{table}[!htb]
	\centering
  \caption{Comparison of the material budget of multi-wire chambers and SciFi profile monitors produced with fibres of \SI{0.5}{mm} or \SI{1}{mm} thickness.}
    \label{tab:results_material}
	\begin{tabular}{lc}
		\toprule
        {Detector} &  {$x/X_0$ (\%)} \\
		\midrule
        MWPC & 0.36\\
        DWC & 0.28\\
        SciFi \SI{0.5}{mm} & 0.24\\
        SciFi \SI{1}{mm} & 0.48\\
		\bottomrule
	\end{tabular}
\end{table}

The main limitations of scintillating fibres are their difficult readout due to the low light signal reaching the photodetectors and the large number of channels that are typically needed.
Nevertheless, the recent invention of the Silicon Photomultiplier (SiPM) \cite{buzhan2003silicon,buzhan2002advanced,dolgoshein2006status} has helped to overcome these limitations.
This new device is a compact solid-state photodetector with high detection efficiency, single-photon detection capabilities, and a low price per channel.
Its major drawback is a high dark count rate, which in certain models can reach levels of \si{MHz/mm^2}

Scintillating fibres are used extensively as beam hodoscopes and tracker detectors in high-energy physics experiments due to their good space and time resolution, high detection efficiency and high-rate capabilities.
Major high-energy physics experiments in the 1980's, such as UA2 \cite{ua2} and D\text{\O} \cite{d0}, were equipped with SciFi trackers that mainly used Charged-Coupled Devices (CCD) to read-out the light generated by the fibres.
In 1991, the FAROS Collaboration (RD-17) at CERN \cite{agoritsas1998development} started an extensive research and development program on new beam hodoscopes based on fine scintillating fibres read-out with Position-Sensitive Photomultipliers (PSPM).
This program led to some important beam hodoscopes, a good example of which is the SciFi hodoscope of the COMPASS Experiment (NA58) at CERN \cite{horikawa2004development} in the early 2000's.
The SciFi tracker stations of COMPASS evolved over time, fostering a rich research and development program \cite{bisplinghoff2002scintillating,gorin2006high,horikawa2002scintillating}.
The standard SciFi station was made of 7 layers of staggered fibres SCSF-78MJ from Kuraray, with circular cross-section and a diameter of \SI{0.5}{mm}.
By using PSPM read-out with specially designed peak-sensing electronics, the COMPASS hodoscope reported (\cite{horikawa2004development}) a spatial resolution of $\sim\SI{125}{\micro\meter}$, a time resolution of $\sim\SI{540}{ps}$ (r.m.s.), and a detection efficiency above 98\% for beam fluxes up to $\sim10^8$ muons per second.

The latest generation of SciFi hodoscopes and trackers have replaced PSPM by newer high-performance Multi-Anode Photomultipliers, for example in the ATLAS ALFA Experiment \cite{green2007design}, or by Silicon Photomultipliers in the Mu3e Experiment \cite{bravar2017scintillating} and the LHCb SciFi tracker \cite{lhcb}. 

\section{Description of the monitor}
\subsection{eXperimental Beam Profile Fibre (XBPF) monitor}
The plastic scintillating fibres used are the SCSF-78 from Kuraray with square cross-section and \SI{1}{mm} thickness.
Other models from the manufacturer Saint-Gobain were also studied (BCF-12 multi-clad), but the SCSF-78 were favoured because of their lower self-attenuation and higher finish quality, as described in \cite{joram_scifi}.
These fibres are composed of a large scintillating core (96\% of the fibre thickness) and a thin cladding that allows trapping part of the scintillation light by total internal reflection.
The core is made of polystyrene mixed with a proprietary formulation of fluorescent dopants that optimise the spectral emission of the fibre at \SI{420}{nm}, matching therefore the quantum efficiency of SiPMs.
Further information on plastic scintillating fibres and their physics principles can be found in \cite{white,leutz}.

The detection area of the XBPF is formed by 192 fibres that are packed along one plane as a single layer (the number 192 optimises the use of the readout electronics, as explained in \cref{sec:beam_profile}).
The square shape of the fibres has two advantages: it allows for an optimal packing and ensures that all beam particles deposit similar amounts of energy in the monitor, therefore homogenising its response.
The fibres are held in place by a structure made of black Polyoxymethylene (POM) -- a thermoplastic frequently used in precision parts -- and two rods of stainless steel that provide the necessary rigidity to the assembly. 
An ultra-thin foil of Kapton\textsuperscript{\copyright}  polyimide of \SI{25}{\micro\metre} thickness is glued over the fibres to maintain their position.

The light from every fibre is read-out on one end by an individual SiPM, indicating which fibre has been activated and thus permitting the reconstruction of the beam profile or the track of a particle from multiple monitors. 
After studying several SiPM from different manufacturers (Hamamatsu, Ketek, and SensL), the model chosen is the S13360-1350 from Hamamatsu.
This model has shown the lowest dark count rate (below \SI{100}{kHz/mm^2}), the lowest cross-talk (below 1\%), and it has a slightly larger active area (\SI{1.3 x 1.3}{mm} against \SI{1 x 1}{mm}), which benefits the optical coupling with the fibres.
A thin mirror foil is glued on the non read-out end of the fibres with the purpose of reflecting back part of the scintillation light, thus increasing the total signal reaching the photomultipliers.

As previously explained, the XBPF has been designed to be vacuum compatible.
The photomultipliers are located outside vacuum, with the fibres exiting via a feed-through that is sealed with an epoxy resin that ensures the required leak tightness and outgassing levels for the CERN Experimental Areas (vacuum level of \SI{e-3}{mbar}).

\begin{figure}[!htb]
   \centering
   \includegraphics*[width=\columnwidth]{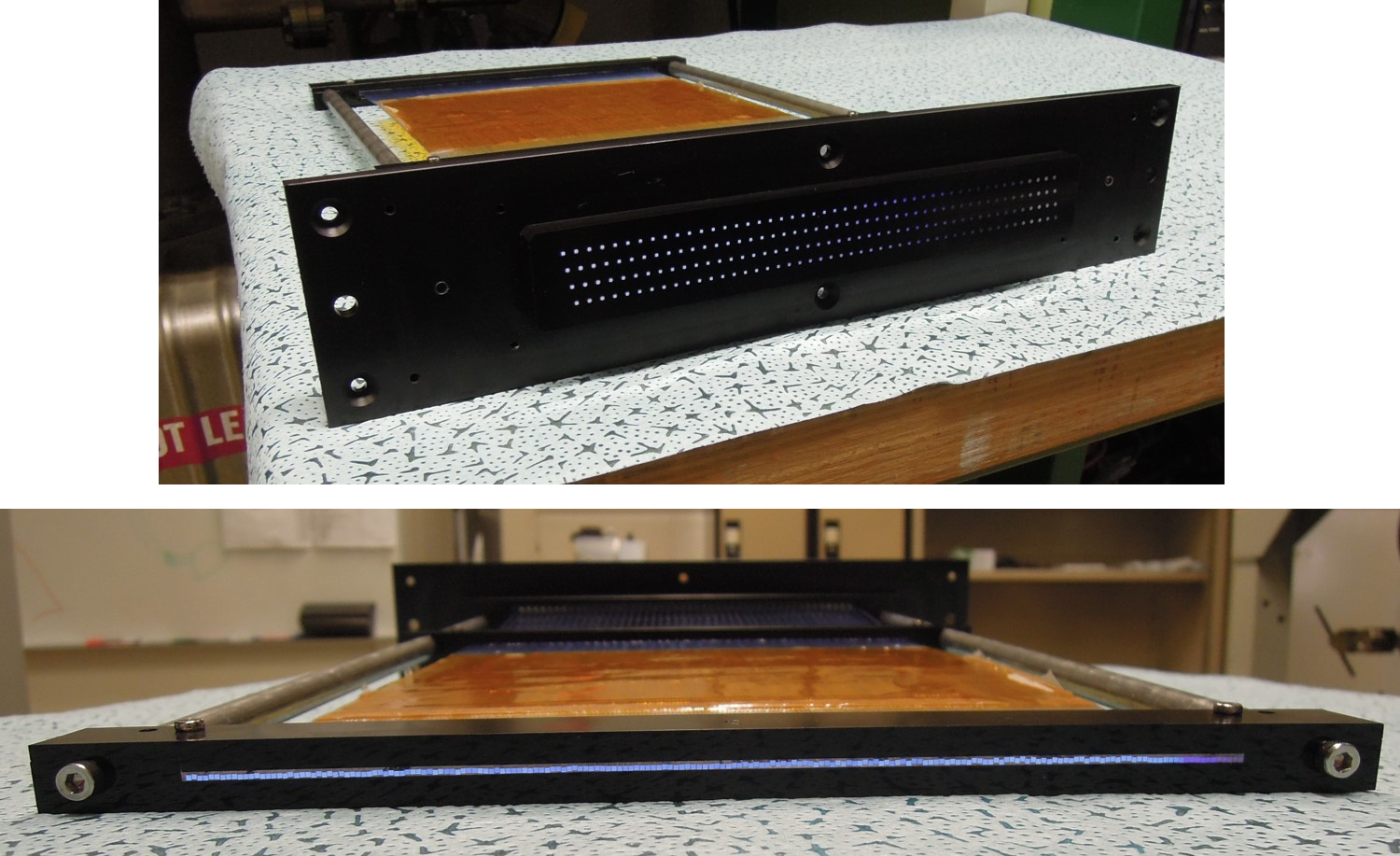}
   \caption{XBPF during assembly. The read-out side and the mirror side of the fibres are shown in detail (top and bottom images, respectively).}
   \label{fig:xbpf_lab}
\end{figure}

\begin{figure}[!htb]
   \centering
   \includegraphics*[width=0.9\columnwidth]{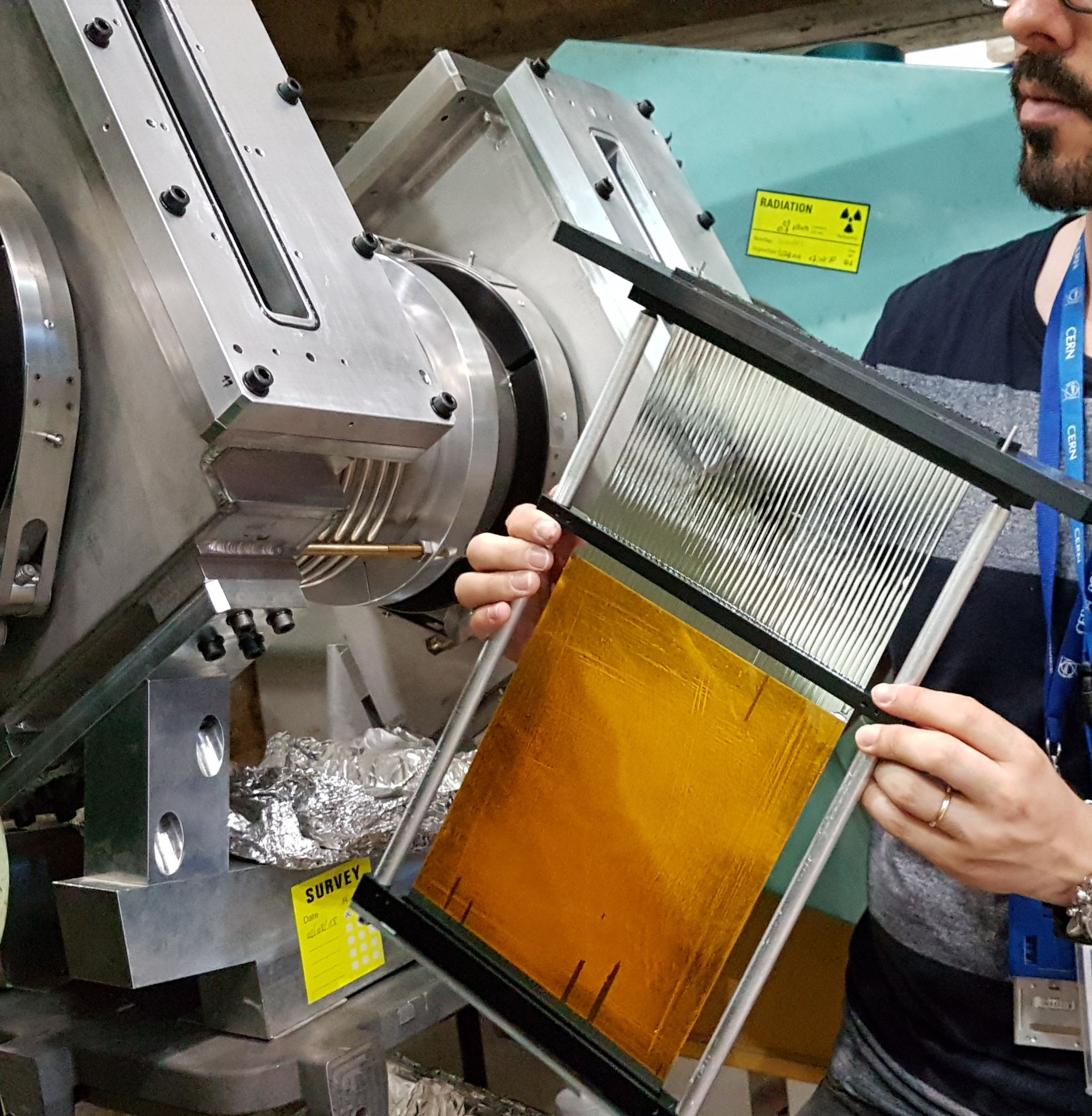}
   \caption{Installation of a XBPF in H4-VLE. The vacuum tank that houses the detector can be appreciated on the left side of the picture.}
   \label{fig:xbpf_install}
\end{figure}

\Cref{fig:xbpf_lab} shows a XBPF during production.
\Cref{fig:xbpf_install} shows a XBPF during installation in H4-VLE.

The fibres of the XBPF have been treated to avoid optical cross-talk between fibres, which is produced by scintillation photons emitted in the ultra-violet that can escape a fibre and excite neighbouring fibres.
The treatment is an ultra-thin vaporisation of $\sim\SI{100}{\nano\metre}$ of aluminium over the fibres.

\subsection{eXperimental Beam Trigger Fibre (XBTF) monitor}
\begin{figure}[!htb]
   \centering
   \includegraphics*[width=0.9\columnwidth]{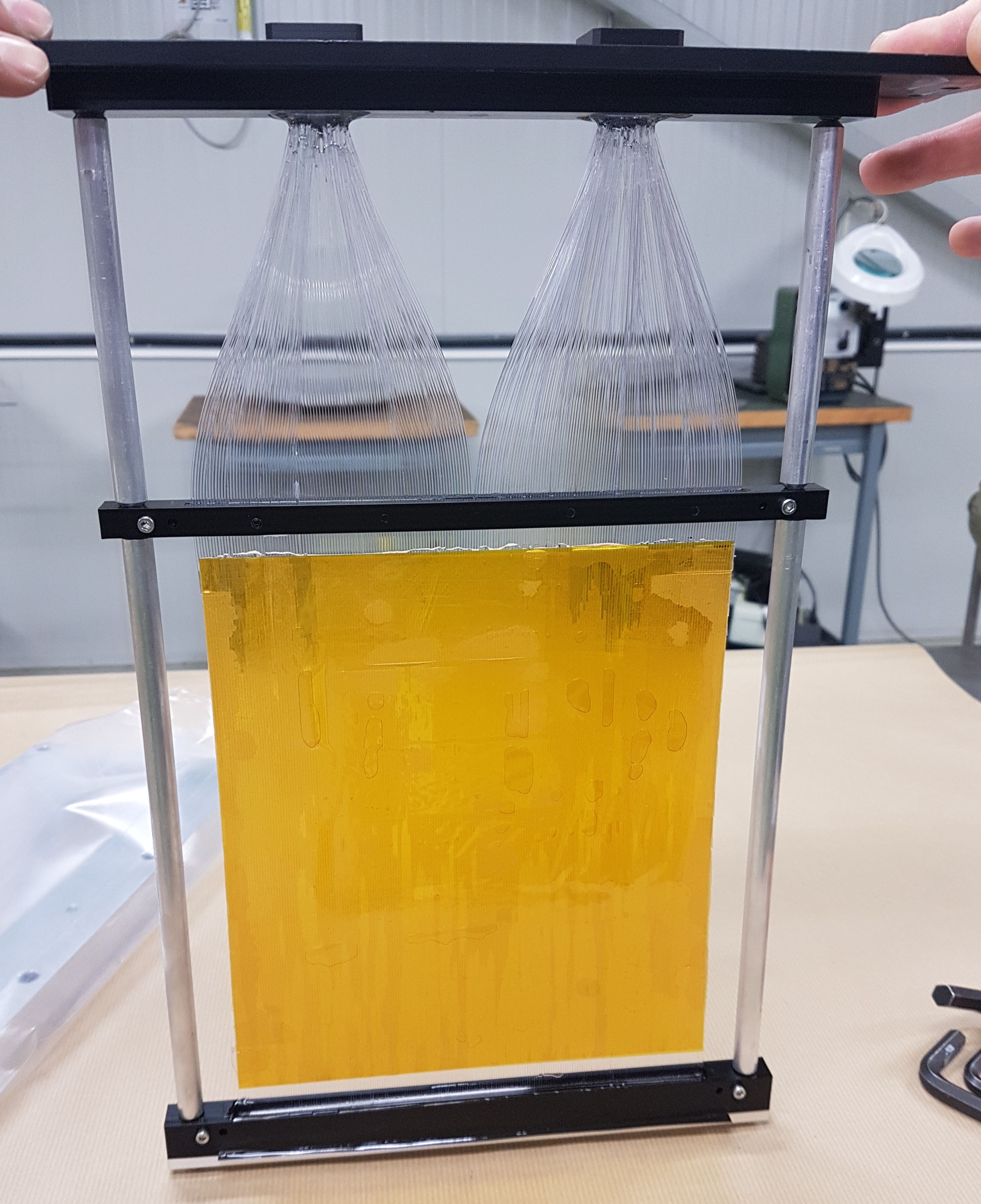}
   \caption{XBTF in the laboratory. In this version of the monitor the fibres are grouped together into two bundles.}
   \label{fig:xbtf_lab}
\end{figure}

A second version of the monitor has been specifically created for the Neutrino Platform with the aim of measuring the beam intensity with high accuracy, of producing fast trigger signals, and of measuring the time-of-flight of the beam particles for their identification.
In this version, the 192 fibres are grouped together into two bundles to be read-out by two Photo Multiplier Tubes (PMT) H11934-200 from Hamamatsu.
These PMT have a low dark count rate of a few counts per second and a low Transit-Time Spread (TTS) of $\sim\SI{300}{ps}$, making them well-suited for TOF applications.
\Cref{fig:xbtf_lab} shows a XBTF ready for testing in the laboratory.

\section{Readout Electronics}
\begin{figure}[!htb]
   \centering
   \includegraphics*[width=\columnwidth]{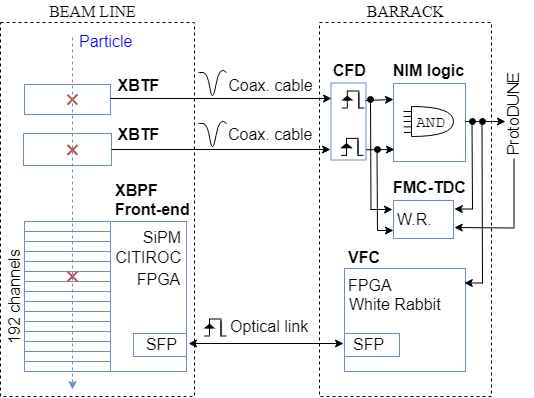}
   \caption{Electronics architecture of the XBPF and XBTF systems.}
   \label{fig:electronics_architecture}
\end{figure}

The electronics architecture of the XBPF and XBTF systems is shown in \cref{fig:electronics_architecture}.
It is divided into three main systems: trigger, beam profile, and timing.

\subsection{Trigger System}
The trigger system combines the signals from several XBTF, creating an unambiguous signal when a beam particle has reached a ProtoDUNE detector.
This beam-trigger signal is sent to the XBPF profile monitors and the ProtoDUNE Experiments to trigger their acquisition.
In the present configuration, the analogue signals from the PMTs are processed by a Constant Fraction Discriminator (CFD) -- the N842 from CAEN.
This type of discriminator is well-suited for timing measurements, as they have a low time jitter and show immunity to signal walk \cite{gedcke1968design}.

\subsection{Beam Profile} {\label{sec:beam_profile}}
The readout electronics of the XBPF profile monitors is formed by a front-end board attached to the detector and a back-end board grouped in a central acquisition chassis in the electronics barrack. 
Both boards communicate via a high-speed optical link.

\begin{figure*}[!htb]
   \centering
   \includegraphics*[width=0.9\textwidth]{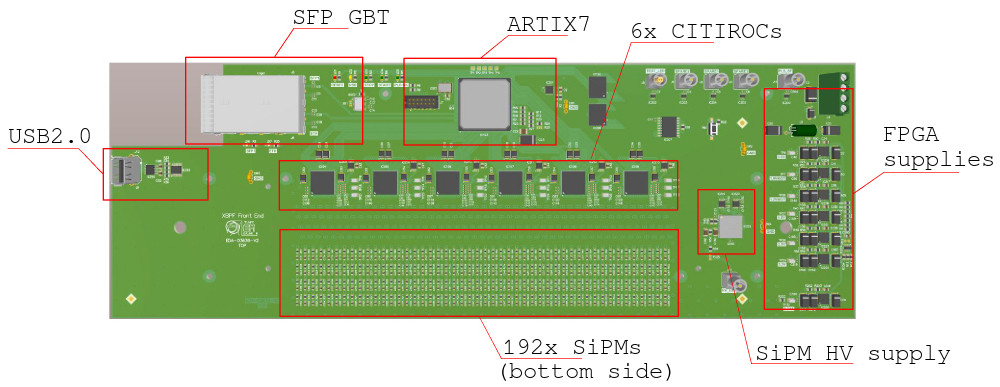}
   \caption{CAD model of the XBPF front-end with its main components highlighted.}
   \label{fig:xbpf_pcb}
\end{figure*}

The front-end board has the following main components, which are highlighted in \cref{fig:xbpf_pcb}:
\begin{itemize}
  \item{192 SiPM that detect the light generated by the fibres.}
  \item{Hamamatsu C11204 to power up the SiPMs. This device has a temperature feedback system to maintain a stable gain of the SiPM.}
  \item{6 CITIROC ASIC \cite{citiroc} that process in parallel the analogue signals from the SiPM.}
  \item{Xilinx FPGA Artix 7 that configures the CITIROC slow control, reads the CITIROC digital output, packages the data, and sends it out in a \SI{10}{MHz} data stream to a Gbit transceiver.}
  \item{A SFP module with Gbit transceiver to transfer the data via optical fibre to the back-end.}
\end{itemize}

The back-end module is the VFC-HD \cite{vfc}, a Versa Module Eurocard (VME) general-purpose digital acquisition board, developed by the CERN Beam Instrumentation group and fully compatible with White Rabbit.
The data stream from the front-end includes both the information from real particles and the noise from the SiPM, which can be at a rate of several kHz when considering all 192 channels.
In order to suppress the noise events, the VFC also receives the global-trigger signal and only records the events coinciding with that signal.
Every recorded event has the information of the status of the 192 fibres (hit, no-hit), plus a White Rabbit \SI{8}{ns}-precision timestamp of when the event occurred.

The XBPF provides simultaneously single particle information and a beam profile integrated over the duration of a burst.
However, single particle information can be switched off when only the beam profile is wanted in order to minimise data generation.

Single particle tracking has been tested in the East Area at CERN up to beam intensities of $\sim 10^5$ particles per second per $mm^2$ \cite{ortega2019multipurpose}.
With the hardware as it is, single particle tracking could be used, in theory, with beam intensities close to $10^7$ particles per second per $mm^2$.
This estimation is made from the sampling rate of the XBPF, which at present is \SI{10}{MHz} per channel.
However, the response of the system at high intensities might depend greatly on the beam characteristics, particularly the time distribution of the particles in the beam.

If only the beam profile and position information is going to be used, for example for beam steering, the maximum limit of $10^7$ could be in principle avoided by reducing the efficiency of the detector in an homogeneous way among all channels.
This could be achieved, for example, by introducing a light filter between the fibres and the SiPM in order to reduce the number of detected photons without altering the shape of the beam profile.

\subsection{Timing system}
This system timestamps digital signals with a resolution of 81 picoseconds in a clock distributed over White Rabbit.
It is based on the FMC-TDC \cite{fmc-tdc}, which is a Time-to-Digital Converter (TDC) board developed at CERN for timing applications using the White Rabbit technology.
This precise timing system is used to timestamp signals from the XBTF, other beam instrumentation, and several signals from the ProtoDUNE trigger system.
It allows the Neutrino Platform physicists to correlate in time the information generated by the beam instrumentation and the neutrino detectors.

Due to the good time resolution of the FMC-TDC, the fast signals of the XBTF are also used to measure the time-of-flight of particles between the first and last XBTFs of the beam line.
The first XBTF is very close (less than \SI{7}{m}) to the beam target where the secondary particles are created, being therefore subjected to a large flux of secondary particles -- around $10^6$ particles per burst.
However, less than $10^3$ particles are captured by the beam line optics and reach the last XBTF, located right in front of the ProtoDUNE Experiment.
In order to avoid fake signals in the TOF system, the beam-trigger signal is also sent to the FMC-TDC board with the aim of filtering the TOF data by selecting only time-matched beam particles.


\section{The H4-VLE Beam Line}
\begin{figure*}[!htb]
   \centering
   \includegraphics*[width=0.7\textwidth]{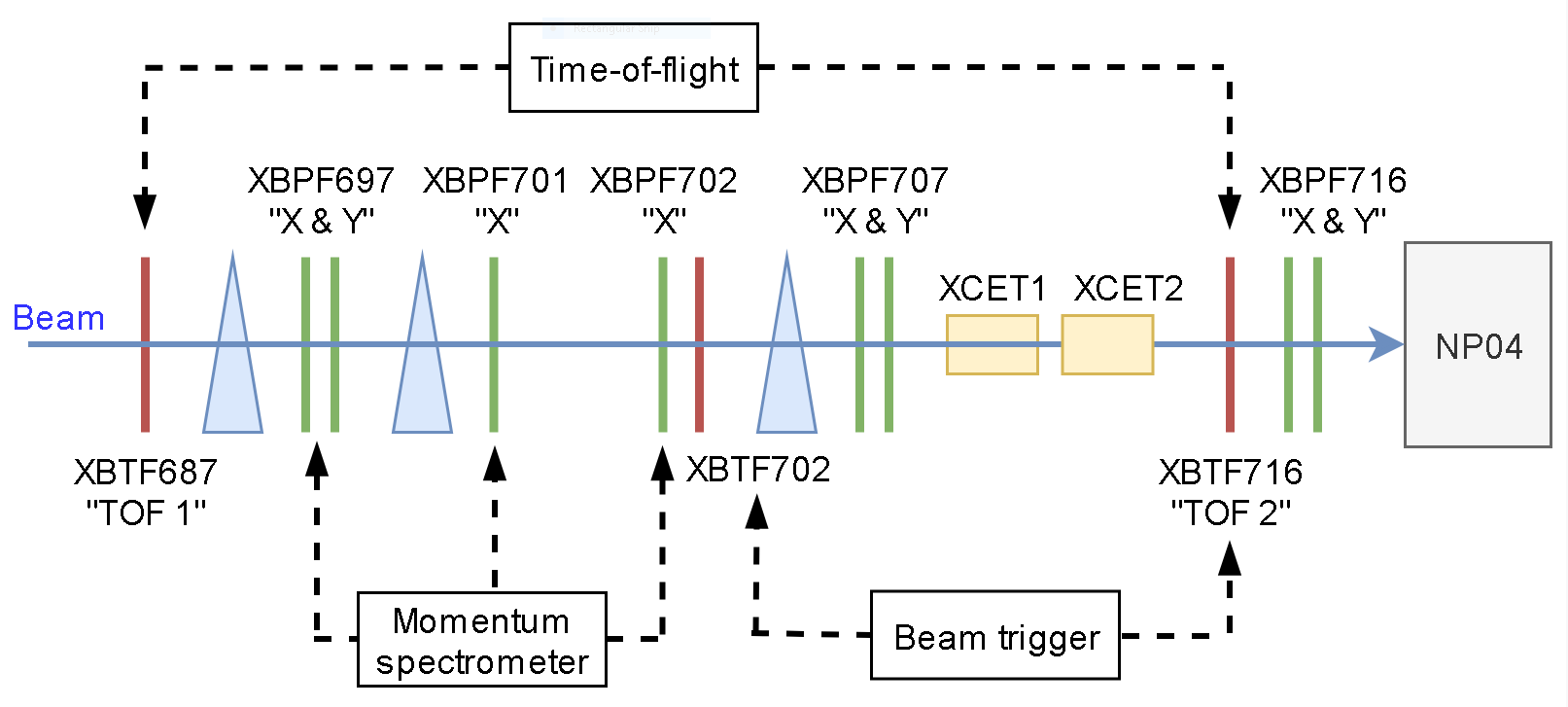}
   \caption{Layout of the H4-VLE beam line indicating the position and function of the different beam monitors. The XBPF are shown in green and the XBTF in red. The main dipole magnets that determine the beam attributes are also shown as blue triangles. The numbers indicate the distance in metres to the primary beam target. Image taken from \cite{h4-vle_paper}.}
   \label{fig:layout_H4}
\end{figure*}

\begin{figure*}[!htb]
   \centering
   \includegraphics*[width=0.9\textwidth]{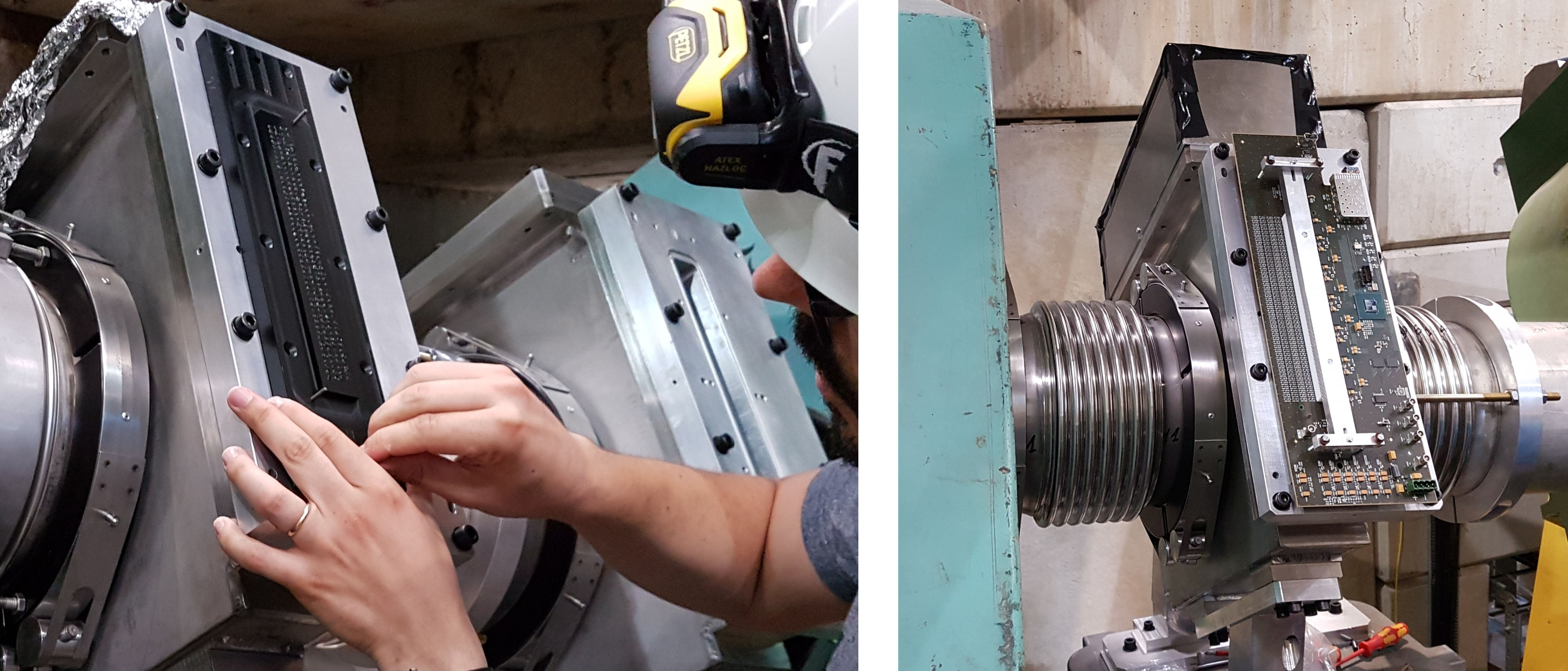}
   \caption{Left: installation of a XBPF fibre plane in a vacuum tank. Right: front-end board placed on top of a fibre plane. The ensemble is afterwards covered with a light-tight box as seen on the back of the picture.}
   \label{fig:xbpf_H4}
\end{figure*}

This paper focuses on the H4-VLE beam line, where a total of 8 XBPF and 3 XBTF have been installed and all systems are fully operational -- profile, trigger, and TOF.
A vacuum tank of the XBPF can accommodate two monitors -- XBPF or XBTF -- in any desired combination, therefore giving great flexibility to the design of the beam line.
The layout of H4-VLE is shown in \cref{fig:layout_H4}, indicating the position and function of every monitor.
\Cref{fig:xbpf_H4} shows two photographs taken during the installation of the monitors.

\begin{figure}[!htb]
   \centering
   \includegraphics*[width=0.9\columnwidth]{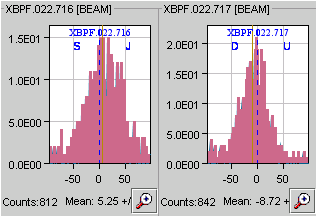}
   \caption{Profile of the first beam successfully delivered to NP04.}
   \label{fig:first_beam}
\end{figure}

Both the XBPF and the H4-VLE beam line were successfully commissioned during the first two weeks of September 2018.
\Cref{fig:first_beam} shows the profile of the very first beam delivered to the NP04 detector, as measured by the XBPFs placed directly in front of the experiment.
The data taking of NP04 took place between the beginning of October 2018 and the middle of November 2018.
During that period, H4-VLE provided protons, pions, kaons, muons, and positrons in the range of momenta \SIrange{1}{7}{GeV/c}, with the addition of two extra runs at sub-GeV range, \SI{0.5}{GeV/c} and \SI{0.3}{GeV/c}.
The maximum and minimum intensities achieved were respectively $\sim1,500$ and $\sim100$ particles per burst.

\subsection{Momentum Spectrometer}
It is possible to measure the momentum of the beam particles with a system composed of three profile monitors around a dipole magnet.
In this configuration, the momentum is calculated from the deflection angle exerted by the magnet to the particle, the length of the magnet, and its magnetic field.
Such a momentum spectrometer has been set up with the "X" planes of XBPF 697 (the number indicates the distance in metres to the primary beam target), XBPF 701, XBPF 702, and a dipole magnet placed between XBPFs 697 and 701.
\Cref{fig:layout_H4} shows a diagram with the layout of the spectrometer.
Further details about the design and operational principle of this spectrometer can be found in \cite{h2}, section 4.1. 

The three XBPF are aligned with the magnetic field of the dipole in such a way that the deflection angle exerted by the magnet can be measured, thus measuring the particles' momentum. 
The length of the magnet is well known and its magnetic field is set up by the beam line physicists according to the desired beam momentum.
In fact, this system acts as a beam momentum station, selecting particles within a certain momentum range which are further carried by the beam line optics to the ProtoDUNE detector.

The performance of the momentum spectrometer during the 2018 data taking period is described in detail in \cite{h4-vle_paper}, section III-D.

\subsection{Cherenkov Threshold Counters}
H4-VLE is also equipped with two Threshold Cherenkov Counters (XCET1 and XCET2), which is a standard beam monitor at CERN for particle identification.
The information of the XCET and TOF systems can be combined (time information and particle-tagging information) to provide a powerful tool for the diagnosis of the beam composition, as presented in \cref{sec:tof}.
More information on the XCET used in the Neutrino Platform can be found in \cite{charitonidis2017estimation}.

\section{Results: Performance of the XBPF and XBTF}
The performances of the XBPF and XBTF systems were excellent over the whole data taking period, in addition to not suffering any downtime.
Furthermore, the data from the monitors are in good agreement with the expected values obtained with Monte Carlo simulations of the beam line. 
The performance of the beam line and its instrumentation has been analysed in detail by a team of beam line physicists of CERN and the Neutrino Platform, who have reported their results in \cite{h4-vle_paper}. 

\subsection{XBPF}
\begin{figure}[!htb]
   \centering
   \includegraphics*[width=\columnwidth]{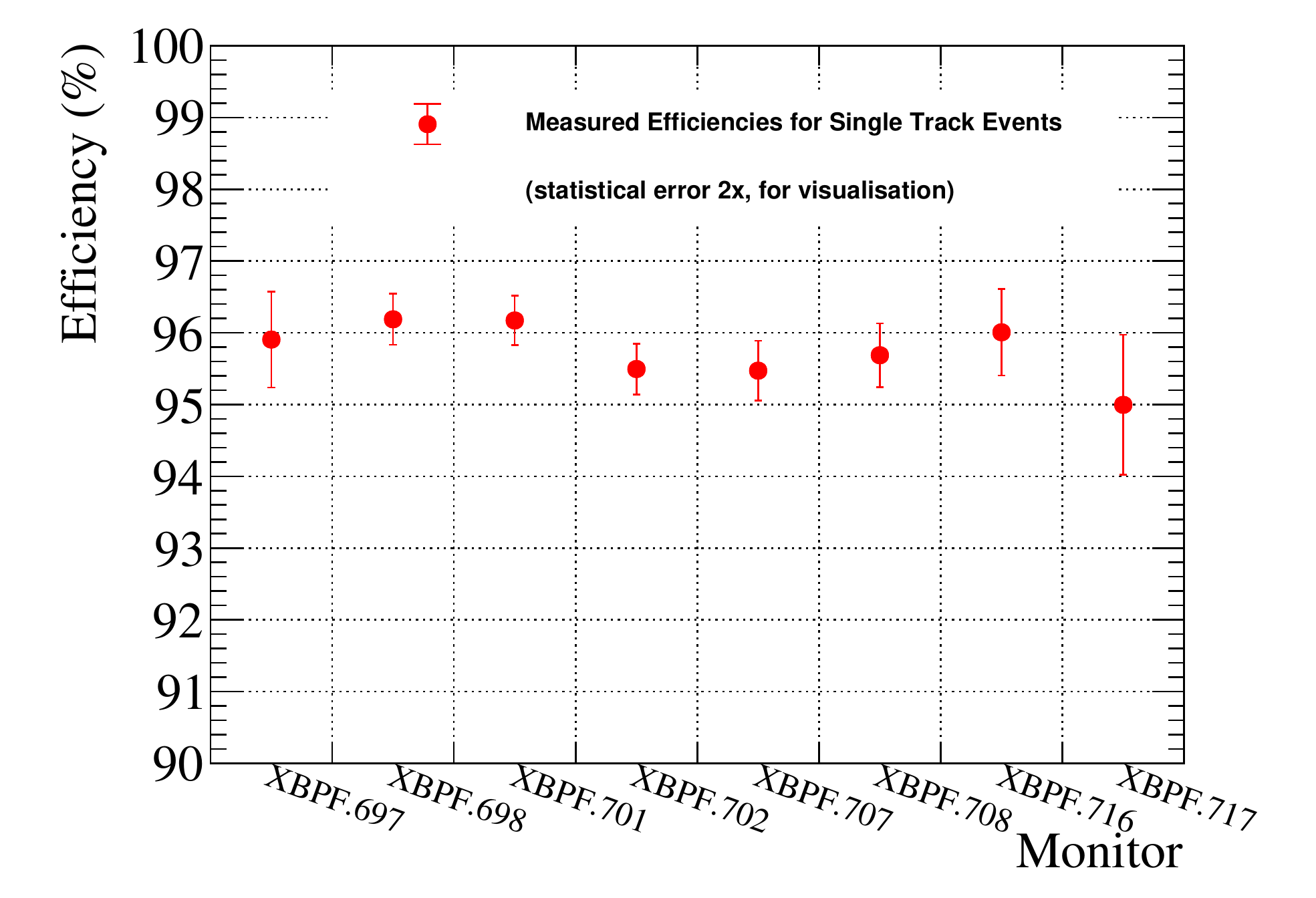}
   \caption{Detection efficiency of the XBPF in H4-VLE for single track events measured during the 2018 data taking. Image inspired from \cite{h4-vle_paper}.}
   \label{fig:xbpf_efficiency}
\end{figure}

The detection efficiency of the XBPF monitors can be easily measured from the number of triggered acquisitions with no fibres activated.
\Cref{fig:xbpf_efficiency} shows the efficiency measured for all XBPF stations during the 2018 data taking.

An offline cut on the hit map has been applied to select exclusively single-track events because we suspect that the first monitors could have an overestimated efficiency value.
These monitors were subjected to very large particle fluxes --higher than $10^6$ particles per second-- due to their proximity to the beam target, therefore exceeding the original specification of 1,000 particles per second (\cref{sec:introduction}).
The XBPFs of the Neutrino Platform were configured to record all fibre hits occurring \SI{500}{ns} before and after a beam trigger signal was received.
This \SI{1}{\micro\second} time window facilitated the alignment of the beam trigger with the XBPF events and seemed safe according to the original specifications of 1,000 particles per second.
However, the unexpected large particle flux may have resulted in several particles being recorded for a given trigger, which would artificially increase the measured efficiency of the detectors.

This issue can be easily solved by reducing the acquisition time window, which does not require any major hardware modification and will be deployed in the next run.

\subsection{XBTF}
\begin{figure}[!htb]
   \centering
   \includegraphics*[width=\columnwidth]{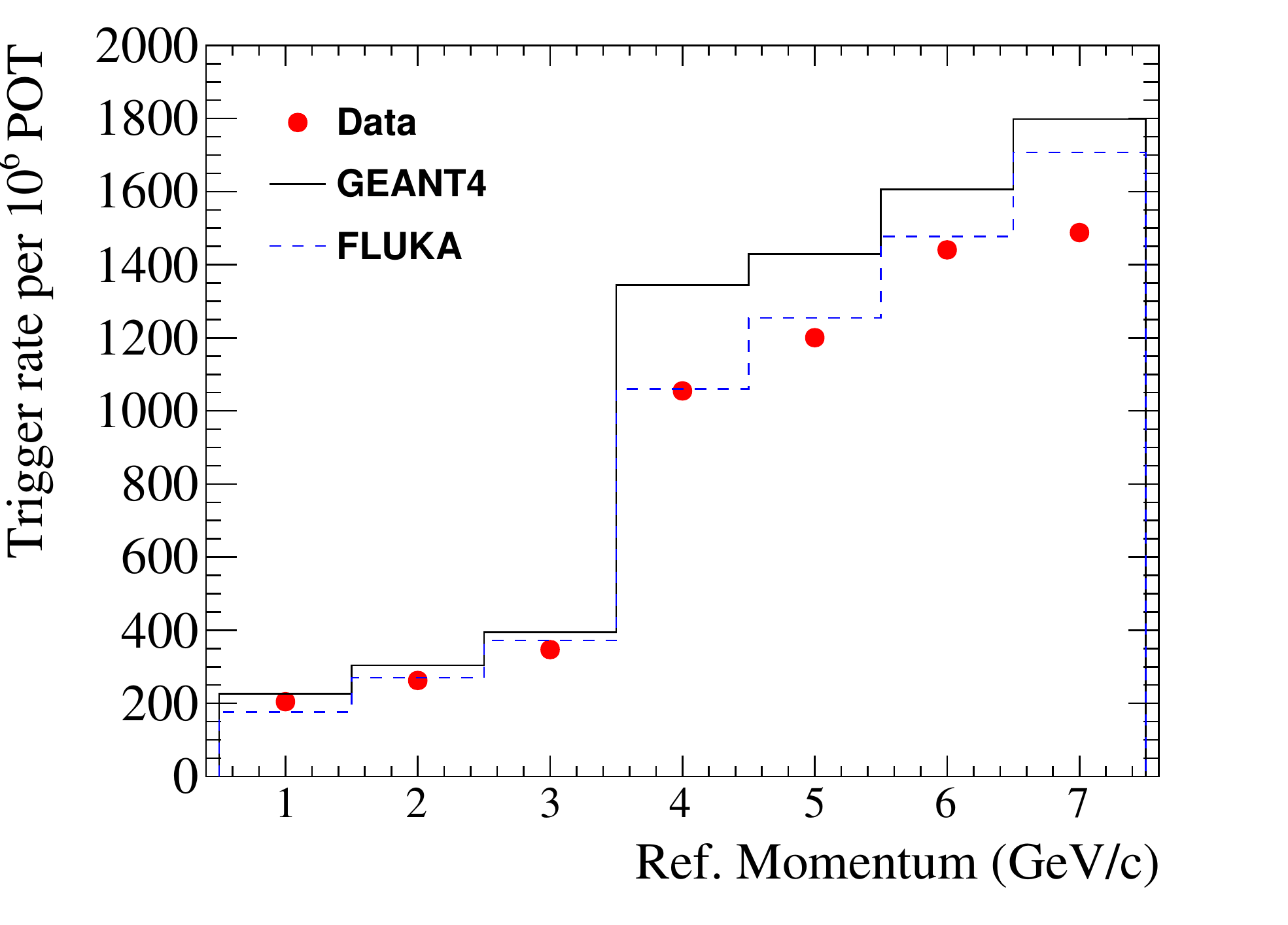}
   \caption{Number of beam-triggers per $10^6$ Particles-On-Target (POT). Comparison between the XBTF measurements and the Monte Carlo simulations of the beam line. An efficiency of 95\% per XBTF trigger plane is assumed in the simulations. Image taken from \cite{h4-vle_paper}.}
   \label{fig:trigger_rate_comparison}
\end{figure}

The efficiency of the XBTF could not be directly measured in H4-VLE, as in the case of the XBPF.
Nevertheless, measurements in a dedicated test bench in the laboratory have shown an efficiency of:
\begin{equation}
  \varepsilon_{XBTF} = 94.0 \pm 0.1 \%
\end{equation}
Such a value is in good agreement with the Monte Carlo simulations of the beam line, as shown in \cref{fig:trigger_rate_comparison}.

%
%

\subsection{Time-Of-Flight} 
\begin{figure}[!htb]
   \centering
   \includegraphics*[width=\columnwidth]{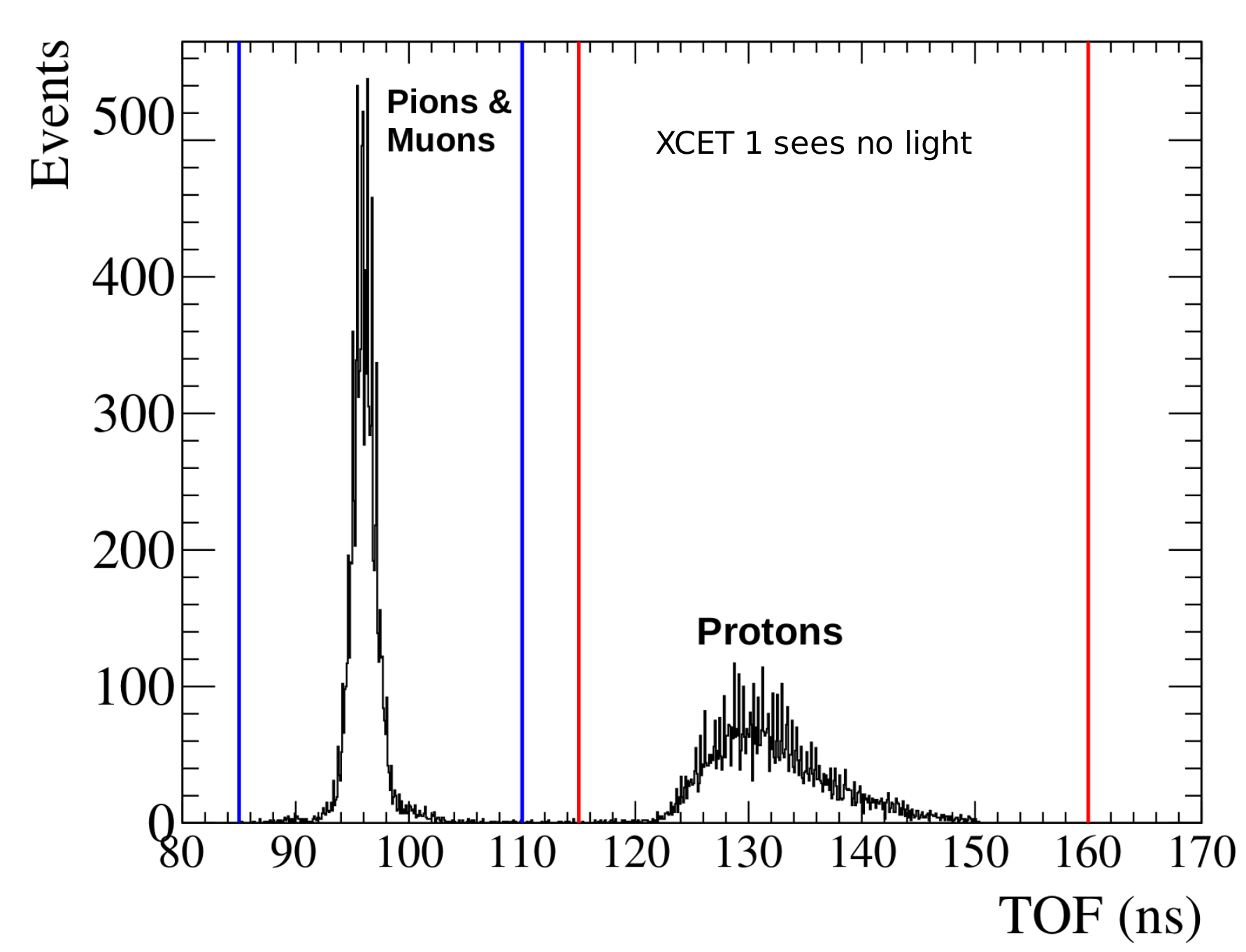}
   \caption{Time-of-flight data combined with the information from the Cherenkov Threshold Counters. At \SI{1}{GeV/c}, when the XCET 1 does not give signal, pions and muons can be identified thanks to the TOF. When XCET 1 gives signal, positrons can be tagged instead. The combination of the two systems provides a powerful tool for the study of the beam composition. Image taken from \cite{h4-vle_paper}.}
   \label{fig:tof}
\end{figure}

{\label{sec:tof}}
The TOF system also performed well, with a measured time resolution of $\sigma_{tof} < \SI{900}{ps}$ (r.m.s.), as expected from previous beam tests (\cite{ortega2019multipurpose}).
Given the \SI{28.6}{m} distance between the two XBTFs of the TOF system, this time resolution makes it possible to differentiate protons from pions, electrons, muons and kaons at momenta below \SI{3}{GeV/c}.
The combination of the TOF and XCET information, as shown in \cref{fig:tof}, is a powerful tool for particle identification that is extensively used by the ProtoDUNE physicists.

We believe that the main factors influencing the time resolution of the TOF system are:
\begin{itemize}
  \item{Generation and propagation of photons inside scintillating fibres.}
  \item{Transit-Time Spread (TTS) of the PMT, which for the H11934 is \SI{300}{ps} according to Hamamatsu.}
  \item{Time walk of the CFD, which for the N842 is \SI{400}{ps} according to CAEN.}
  \item{Conversion of the NIM signals from the CFD to TTL logic level with the module N89 from CAEN. Such conversion is necessary because the FMC-TDC only accepts TTL signals. The jitter introduced by this module has been measured by us to be \SI{110}{ps}.}
  \item{Time resolution of the FMC-TDC, which is \SI{81}{ps} according to our own measurements.}
\end{itemize}

On a first order approximation, it can be assumed that most of these factors are independent and can be summed in quadrature \cite{conti2011improving}:
\begin{equation} {\label{eq:tof_propagation}}
  \sigma_{tof}^2 = \sigma_{fibre}^2 + \sigma_{pmt}^2 + \sigma_{cfd}^2 + \sigma_{nim-ttl}^2 + \sigma_{tdc}^2
\end{equation}
which allows us to estimate the contribution from the light generation and propagation within the fibre: $\sigma_{fibre}= \sim \SI{735}{ps}$.

We believe that the main limitations to the performance of the TOF system are:
\begin{itemize}
  \item{The large time walk of the CFD and the conversion to TTL logic level.}
  \item{Photons bouncing back from the mirror of the XBTF that smear the time resolution of the fibres.}
\end{itemize}

It is foreseen to install in the next run a new model of CFD, the 6915 from Phillips Scientific, which features 75ps time walk and TTL signal output. 
Thanks to this upgrade, the time resolution is expected to improve to \SI{800}{ps}, which will facilitate particle identification at \SI{3}{GeV/c}.

Concerning the mirror, unfortunately it is not possible to remove it since that would compromise the efficiency of the trigger system, which is more important for the ProtoDUNE Experiments than a better TOF resolution. 
A possible solution would be to install new XBTFs without mirror exclusively dedicated to TOF in the available space of the present XBTF vacuum tanks. 

A measurement of the light yield of the XBTF would help to better understand the contribution from the scintillating fibres.
Unfortunately, it was not possible to perform such measurement before the commissioning of the Neutrino Platform beam lines due to lack of time but it is programmed for the next run. 

Similar scintillating fibre hodoscopes, such as a prototype of the COMPASS hodoscope, have reported a time resolution of about $\sigma = \SI{400}{ps}$ in dedicated beam tests \cite{bisplinghoff2002scintillating}.
This prototype module was made of three staggered layers of Kuraray SCSF-78 multi-clad fibres with circular cross-section and \SI{1}{mm} diameter, which were read-out by Multi-Anode Photomultipliers (H6568 from Hamamatsu with a TTS of \SI{280}{ps}), measuring a light yield of around 30 photoelectrons per Minimum Ionising Particle.

The more modern experiment Mu3e \cite{bravar2017scintillating} is developing a SciFi tracker for timing measurements that achieves a time resolution $\sigma < \SI{500}{ps}$.
It is based on three staggered layers of square fibres Saint-Gobain BCF-12 of \SI{250}{\micro\meter} with SiPM (Hamamatsu 13360-1350) coupled on both ends and read-out by a dedicated chip, the MuTRIG readout chip \cite{chen2017mutrig}.

\section{Conclusions and Outlook}
The new scintillating fibre monitor has completely fulfilled the requirements of the Neutrino Platform in terms of performance and functionality. 
The XBPF has measured the beam profile and the beam momentum with high efficiency and the required spatial resolution.
The XBTF has measured the beam intensity and provided trigger signals to the NP04 Experiment with an efficiency similar to the XBPF.
Moreover, the signals from the XBTF have been reused to create a time-of-flight system with a time resolution of $\sim\SI{900}{ps}$, which has provided valuable particle identification for the lower momentum beams. 
Thanks to the low material budget of the monitors, the minimum achieved momentum in H4-VLE was \SI{0.3}{GeV/c}, exceeding the original specifications.

New Constant Fraction Discriminators with lower time walk and direct TTL signal output have been tested in the laboratory and will be implemented in the next run.
The time resolution of the TOF is expected to improve to $\sigma = \SI{800}{ps}$ after this upgrade.

At present, the XBPF is being produced for the East Experimental Area at CERN, which is undergoing a complete renovation to be finished by 2021 \cite{bernhard2018beamlines}.
The XBPF will replace all gaseous wire chambers, thus consolidating in this way the profile monitoring in this facility. 

Some studies on radiation hardness of Kuraray SCSF-78 fibres report that up to total absorbed doses of approximately \SI{100}{kGy}, the fibres keep around 80\% of their detection efficiency \cite{mapelli2011scintillation}.
Monte Carlo simulations of the XBPF described in \cite{ortega2018accurate} foresee a life span of several decades in the CERN Experimental Areas before reaching that dose.
However, a comprehensive study on the radiation damage of polystyrene based scintillating fibres \cite{busjan1999shortlived} reports that annealing processes have a major impact in the radiation hardness and such processes need an atmosphere rich in oxygen to occur.
Since the XBPF are placed in vacuum, their radiation hardness might be compromised.
The same study \cite{busjan1999shortlived} reports that a later exposition to oxygen triggers the annealing processes, meaning that perhaps an eventual exposure of the XBPF to air, as happens for example during technical stops, may be enough to recover the fibres from radiation damage.

In case of severe degradation of the performance due to radiation damage, the modular design of the XBPF allows for a simple and cost-effective replacement of the damaged fibres, with a cost lower than 1,000 CHF per fibre plane. 

The present design of the XBPF has been chosen based on a compromise between performance and cost.
The monitor is inexpensive and can be produced by the Beam Instrumentation group at CERN on demand. 
It is easy to maintain, has a low power consumption and does not need gas or cooling systems.
If a certain application requires higher performance, the XBPF can be easily modified to accommodate several layers of staggered fibres in order to improve the detection efficiency and spatial resolution. 
However, such modifications tend, as a trade-off, to raise the overall cost and to complicate the production processes.

\section{Acknowledgements}
The authors would like to thank particularly W. Devauchelle for his invaluable help in the design and development of the monitor and M. Raudonis for his work on different trigger techniques of the XBPF. Similarly, the authors would like to thank A.C. Booth, N. Charitonidis, P. Chatzidaki, L. Gatignon, Y. Karyotakis, E. Nowak, M. Rosenthal, and  P. Sala, for their significant contribution to the design of the XBPF and to the analysis of the data of the monitors in H4-VLE. The whole project has been only possible because of the help and good advice from the following colleagues: M. Azevedo, M. Barros, A. Bay, A. Boccardi, P. Carriere, S. Deschamps, R. Dumps, A. Ebn Rahmoun, S. Excoffier, S. Girod, E. Gousiou, G. Haefeli, M. Hrabia, S. Jakobsen, R. Jones, C. Joram, S. Kaufmann, M. Mclean, M. Ricci, J. Serrano, J. Spanggaard, F. Vaga, and E. Van Der Bij.



\bibliographystyle{elsarticle-num}
\bibliography{bibliography}

\end{document}